\documentclass[aps,pra,twocolumn]{revtex4-1}
\usepackage{amssymb,amsfonts,amsmath,graphicx,color,mathptm,times}
\usepackage{hyperref}
\usepackage{soul,cancel}

\newcommand{\beq}{\begin{equation}}
\newcommand{\eeq}{\end{equation}}
\newcommand{\bea}{\begin{eqnarray}}
\newcommand{\eea}{\end{eqnarray}}

\newcommand{\bra}[1]{\left< #1 \right|}

\newcommand{\ket}[1]{\left| #1 \right>}

\begin{document}
\title{Einstein-Podolsky-Rosen Steering and Quantum Steering Ellipsoids: \\
Optimal Two-Qubit States and Projective Measurements}

\author{R. McCloskey}
\affiliation{Centre for Theoretical Atomic, Molecular, and Optical Physics, School of Mathematics and Physics, Queen's University, Belfast BT7 1NN, United Kingdom}

\author{A. Ferraro}
\affiliation{Centre for Theoretical Atomic, Molecular, and Optical Physics, School of Mathematics and Physics, Queen's University, Belfast BT7 1NN, United Kingdom}
  
\author{M. Paternostro}
\affiliation{Centre for Theoretical Atomic, Molecular, and Optical Physics, School of Mathematics and Physics, Queen's University, Belfast BT7 1NN, United Kingdom}

\date{\today}

\begin{abstract}
We identify the families of states that maximise some recently proposed quantifiers of Einstein-Podolsky-Rosen (EPR) steering and the volume of the Quantum Steering Ellipsoid (QSE). The optimal measurements which maximise genuine EPR steering measures are discussed and we develop a novel way to find them using the QSE. We thus explore the links between genuine EPR steering and the QSE and introduce states that can be the most useful for one-sided device-independent quantum cryptography for a given amount of noise.
\end{abstract}

\pacs{ insert pacs}

\maketitle

\section{Introduction}

Despite dating back to 1935~\cite{schr}, Einstein-Podolski-Rosen (EPR) steering has attracted considerable attention only relatively recently~\cite{others,others2,others3}. This phenomenon entails a form of quantum correlations that lies between entanglement and non-locality: while not all entangled states are steerable, there are steerable resources that do not violate a Bell inequality. It  refers to the possibility (without any classical counterpart) for one agent to remotely change ({\it steer}) the state of another one by performing local measurements on his own subsystem of a suitable entangled resource that the two agents share. The landmark paper by Wiseman, Jones, and Doherty~\cite{Wise07} has opened up the way to the application of the tools of entanglement theory to EPR steering. Several convex monotones have been proposed to date for the quantification of EPR steering~\cite{Miguel,Piani,Gallego}. Moreover, there have been significant advancements in the ideas of temporal and spatio-temporal steering and their connection to non-Markovian open-system dynamics~\cite{Nori}.

From the quantum communication viewpoint, the relevance of steering is remarked by its resource role in quantum key distribution with an untrusted party~\cite{branciard}, which echoes the operational interpretation to such notion given in terms of entanglement distribution by an untrusted party given in~\cite{Wise07}, and which grounds EPR steering as a valuable resource in protocols where only one party is trustworthy.

Notwithstanding the fundamental and technological prominence of EPR steering, its quantification is currently elusive. Among recent proposals for the identification of suitable ways to quantify steering, the one based on quantum steering ellipsoid (QSE)~Ref.~\cite{Sania} is particularly appealing. QSEs provide a useful visualisation of any two-qubit state.  Let us assume that our two-qubit state is shared by Alice and Bob who each have one half of the system in question. The QSE for Alice shows the influence that Bob can have on her qubit by making projective measurements on his qubit and sending the corresponding measurement statistics to Alice. All possible projections that Bob can make correspond to the surface of an ellipsoid on the Bloch sphere for Alice, while positive operator valued measures (POVMs) correspond to points in the interior. Such QSE, together with the two parties' Bloch vectors, give a geometric representation of the shared state. The volume of a QSE, on the other hand, provides a measure of correlations between the qubits. For a Bell state shared by Alice and Bob, the volume of the corresponding QSE is ${4\pi}/{3}$, as this is the only two-qubit state from which Bob's measurements can project Alice's state anywhere. Extensions of this notion to the multi-particle scenario have been recently assessed~\cite{milne,cheng} 

Although the connections between EPR steering and QSEs have been addressed~\cite{Sania2}, little is yet known about the structure of the states that maximize such new correlation measures, much in the spirit of what has already been achieved for entanglement~\cite{Munro} and discord ~\cite{James,Roberta,Mauro}. 

In this paper we fully characterize the distribution of two-qubit states in the purity-vs-volume of the QSE plane, identifying the family of extremal states, i.e. those two-qubit states that maximise the volume of QSE at a set value of purity. We dub such states Maximally Steerable Mixed States (MSMS), in line with the tradition of analogous boundary families maximizing entanglement~\cite{Munro} and discord~\cite{Roberta,James,Mauro}. We find that the frontier of such distribution includes some of the known maximally discordant states at a set value of the shared classical correlations~\cite{Roberta}. Such states are found to be extremal also in the plane consisting of linearised von Neumann entropy and steerable weight~\cite{Miguel}. On the other hand, when the robustness of steering~\cite{Piani} is used, a richer structure of boundary states is found that has no obvious relation with other figures of merit for quantum correlations, thus highlighting the profound differences existing between the latter indicator of steering and geometric measures such as the volume of QSEs.

Then we present a method devised to optimise the amount of steerability obtained from the steerable weight by choosing the measurements by Bob that maximise his effect on Alice. We obtain these optimal projections from the QSE. 
We first address the states that should be used in the presence of white noise and we second consider the optimal projectors which should be used to find the optimum amount of steering.

The remainder of this paper is organised as follows. In Sec.~\ref{tool} we formally introduce the concept of EPR steering and the tool embodied by QSE. Sec.~\ref{optimal} studies the distribution of states in the plane consisting of the volume of QSE or steering weight against the linear entropy, identifying the frontier families for both distributions and comparing them with what is achieved when the robustness of steering is used instead. In Sec.~\ref{basis} we illustrate our proposal for the identification of a quasi-optimal measurement basis for the sake of quantifying the indicators of quantum correlations used in Sec.~\ref{optimal}. Finally, Sec.~\ref{conc} offers our conclusions and presents some questions opened by our investigation. 

\section{EPR Steering and the Volume of the QSE}
\label{tool}

In this Section we aim at recalling the concept of steering and some of the tools to quantify it. Measures of genuine EPR steering are quantifiers which describe to what extent the correlations that are being observed between measurements on Bob's side and the reduced state of the system on Alice's side can be described by a \emph{local hidden variable - local hidden state} (LHV-LHS) model. This is a model which has on Bob's side simply the details of the measurement he performed and the outcome of his experiment, and on Alice's side complete tomographic information on her post-measurement state. This can be described by a joint LHV-LHS model if our information about the system can be written in the form
\begin{equation}
\label{condition}
P(a,b|A,B;W)=\sum_{\xi}P(a|A,\rho_\xi)P(b|B;\xi)P_\xi,
\end{equation}
where $A$ and $B$ are the observables used for the local measurements, each having outcome $a$ and $b$ respectively, $W$ represents the bipartite state under consideration, $P(b|B,\xi)$ and $P_\xi$ are probability distributions involving the LHV $\xi$, and $P(a|A,\rho_\xi)$ is the LHS dependent on it. The state of a bipartite system is considered genuinely EPR steerable if Eq.~(\ref{condition}) does not hold. The first measures of steerability were violations of Bell-like EPR steering inequalities. In this paper we deal with newer measures which involve convex optimisation.

The scenario under scrutiny is that of Alice and Bob sharing a bipartite state. Here, we shall restrict our attention to two-qubit states. Bob performs a (general) measurement $B$ on his subsystem whose outcome $b$ he communicates to Alice. As a result, she builds her {\it assemblage}, i.e. the subnormalised post-measurement state of her qubit 
constructed using the information she received from Bob. Formally
\begin{equation}
\sigma_{b|B}=\{P_{b|B}(b|B),\rho(b|B)\}
\end{equation}
with $\rho(b|B)$ 
the post-measurement states of Alice's qubit each associated with a conditional probability $P_{b|B}(b|B)$. The assemblage is used to describe our state of knowledge about the joint two-qubit system. 

Any two-qubit state can be written in the Pauli basis as $\rho=1/4\sum_{\mu,\nu=0}^{3}\Theta_{\mu\nu}\sigma_\mu\otimes\sigma_\nu$ with $\Theta_{\mu\nu}=\text{Tr}[\rho\sigma_\mu\otimes\sigma_\nu]$, $\sigma_0=\openone$ and $\sigma_{\mu\neq0}$ the $\mu$ Pauli operator (with $\mu=1,2,3$). Written as a Bloch matrix, the $16\times16$ correlation matrix takes the form 
\begin{equation}
\Theta=\left(\begin{array}{cc}
1 & {\bf b}^T\\
\bf a & T
\end{array}\right),
\end{equation}
where $\bf a$ and $\bf b$ are the Bloch vectors of the reduced state of Alice and Bob respectively, and $T$ is a correlation matrix. 
The QSE is at its most comprehensible when one considers the case where  $b=0$. Here, any state ${\bf y}$ that Bob can steer Alice to is given by ${\bf y}={\bf  a} +T {\bf x}$ where ${\bf x}$ is the unit sphere (the Bloch sphere for Alice) shaped and rotated by $T$, centered at $ {\bf a}$. In the case where $b\neq0$ we can transform the full two-qubit state via stocastic local operations aided by classical communication (SLOCC) into $\rho'=(S_A\otimes S_B) \rho(S_A\otimes S_B)$ in order to center $\bf b$. To find our corresponding parametrisation we should find
$$\tilde{\Theta}=\left(\begin{array}{cc}
1 & {\bf 0}^T\\
{\tilde{\bf a} }& \tilde{T}\end{array}\right).$$
Following Ref.~\cite{Sania}, we define the steering ellipsoid as the set of states that Bob may steer Alice to. Any state ${\bf y}$ that Bob can steer Alice to must be given by ${\bf y}={\bf \tilde a} +\tilde{T} {\bf x}$ where ${\bf x}$ is the unit sphere shaped and rotated by $\tilde T$, centered at $\tilde {\bf a}$.
The volume of Alice's QSE is given by
\begin{equation}\label{eqvol}
V_A=\frac{64\pi}{3}\frac{|\text{det}\rho-\text{det}\rho^{T_B}|}{(1-b^2)^2},
\end{equation}
where $b=|{\bf b}|$ is the length of Bob's Bloch vector and and $\rho^{T_B}$ stands for the partial transposition of state $\rho$ with respect to Bob's qubit.

The volume of a QSE make no considerations about what constitutes classical or quantum correlations. Therefore a non-zero QSE volume does not imply that we have an EPR steerable state, as the correlations could be described classically. Notwithstanding this, the notion of QSE is conceptually related to EPR steering in that it shows the effect that Bob can have on Alice through his measurements. In the remainder of this paper, when referring to steering, steerable or steerability, we will be specifically onsidering the special case of genuine EPR steering. We will show that the relation between genuine EPR steering and the volume of QSEs is more than just a conceptual one, and that some quantifiers of the former behave similarly to the latter, when assessed against the families of states that maximize them at set degree of mixedness.

\section{Extremal states in the entropy plane}
\label{optimal}

Here we investigate the families of two-qubit states that maximise the measures considered above when we set the degree of mixedness of the states themselves. The latter will be quantified by the linearised entropy, which is defined as~\cite{peters}
\begin{equation}
S_L=\frac43(1-\text{Tr}[\rho^2])
\end{equation}
with $\rho$ the density matrix of the two-qubit state at hand. The linearised entropy embodies a lower bound to the von Neumann entropy of a state, from which it is obtained as a low-order Mercator-series expansion. We have $S_L=0$ for a pure state of two qubits, while $S_L=1$ for a maximally mixed state. 

We now perform a thorough comparison of the three quantifiers recalled above: QSE, steerable weight, and
robustness of steering. We will highlight the features related to each of the latter, providing in particular the
optimal families of states that maximise them.

\subsection{Optimal families for the volume of the QSE}
\label{optimal}
\begin{figure}
\begin{center}
\includegraphics[width=1.\columnwidth]{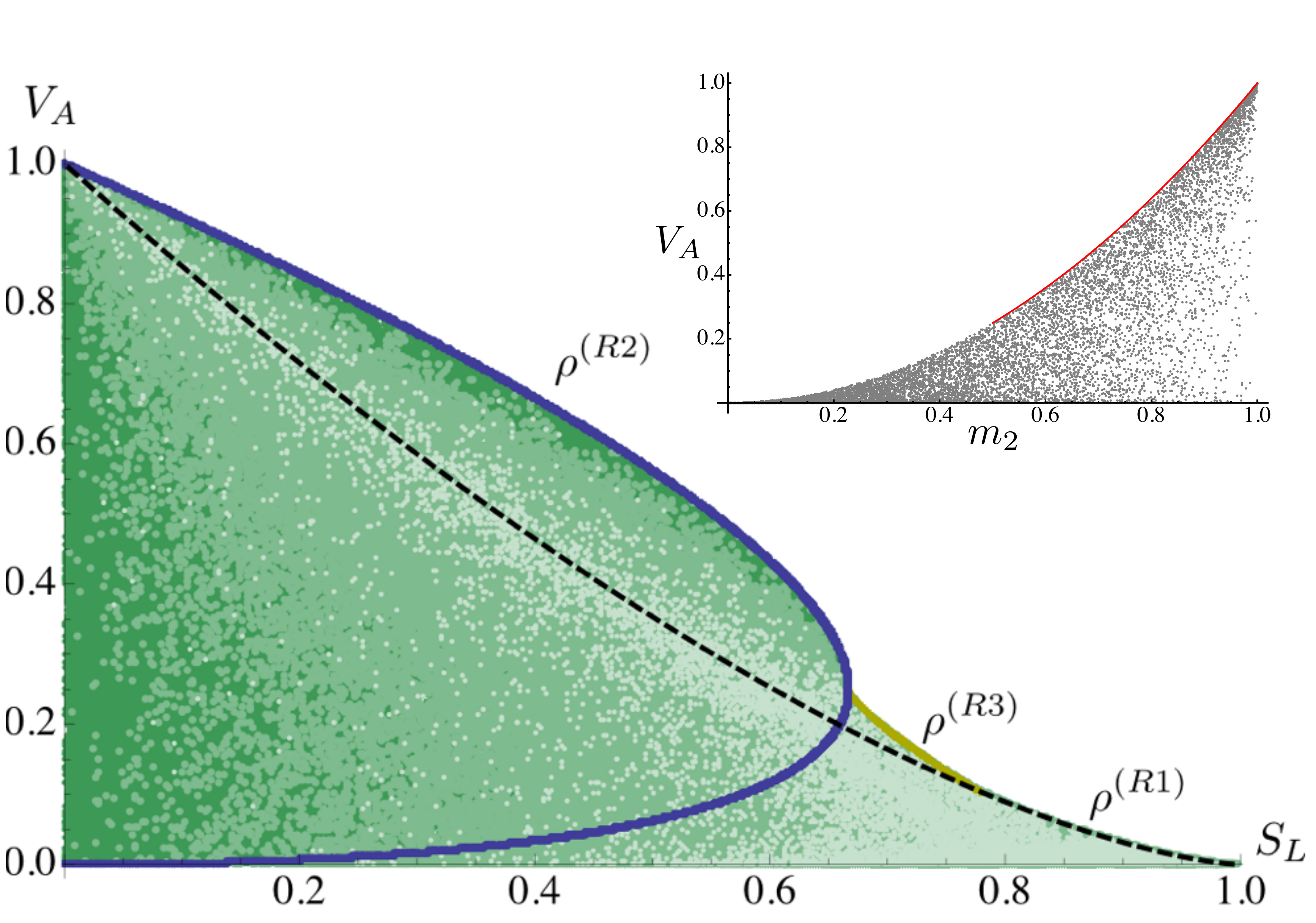}
\caption{(Color online) 
Distribution of the normalised volume of the QSEs (i.e. $V_A/(4\pi/3)$) versus linear entropy for $10^8$ random two-qubit states (green dots). The area of the graph is populated by random states (green), with X-states shown in lighter shades. The three families of states which maximise the volume are shown. The Werner states ($\rho^{(R1)}$, black, dashed) maximise the volume at high linear entropy. The maximally discordant states ($\rho^{(R2)}$, blue, solid) maximise the volume up to a value of ${2}/{3}$ for the linear entropy. The third family $\rho^{(R3)}$, which results from the linear mixture of the first two, is shown in yellow. Inset: we show the functional relation between $V_A$ and $m_2$ for family $\rho^{(R_2)}$ and $m_2\in[1/2,1]$ (which corresponds to the region $S_l\in[0,2/3]$ where $\rho^{(R_2)}$ is maximal). Gray dots are the results of our numerical simulation, while the red curve is the analytic function linking the volume of Alice's QSE to $m_2$.}
\label{figvol}
\end{center}
\end{figure}

We find three optimal families for the QSE volume, the regimes in which they are applicable is shown in Fig.~\ref{figvol}. The first family, $\rho^{(R1)}$, operating in the high linear entropy regime, is given by the Werner states,

\beq
\rho^{(R1)}=m_1\ket{\Phi^+}\bra{\Phi^+}+\frac{1-m_1}{4}\mathcal{I},
\eeq
where $\ket{\Phi^+}=(\ket{00}+\ket{11})/\sqrt{2}$ is one of the maximally entangled Bell states ($\{\ket{0},\ket{1}\}$ is the computational basis), $\mathcal{I}$ is the $4\times4$ identity matrix, and $m_1$ is a mixing parameter. 

The second family of states, $\rho^{(R2)}$ is given by the same set of states found to maximise the discord for a given amount of classical correlation~\cite{Roberta}. The discord is measured thus,

\beq
\delta_{A:B(\rho)}=\min_{\{\Pi_i^B\}}\left[S(\rho_B)-S(\rho)+S(A|\{\Pi_i^B\})\right],
\eeq
where $S$ is the von Neumann entropy, $\rho_B$ is the reduced state of Bob's subsystem and $S(A|\{\Pi_i^B\})$ is a conditional entropy after a projective measurement has been performed on Bob's side. The second family is given by 
\beq
\rho^{(R2)}=m_2\ket{\tilde{\Phi}^+}\bra{\tilde{\Phi}^+}+(1-m_2)\ket{01}\bra{01}
\eeq
with $\ket{\tilde{\Phi}^+}=\sqrt{p}\ket{00}+\sqrt{1-p}\ket{11}$ ($p\in[0,1]$) and $m_2$ the mixing parameter with the pure state $\ket{01}\bra{01}$. This family is valid in the linear entropy range $S_L\in\left[0,2/3\right]$. This is the most interesting region in terms of genuine EPR steerability, since it is unlikely that we will find steering at high values of linear entropy. The blue boundary states in Fig.~\ref{figvol} are in fact given by a subset of $\rho^{(R2)}$ with $p=0.9999$ where $m_2$ controls the position on the line. In fact, $p$ can be taken to be arbitrarily close to $1$, making the correlation terms in the density matrix, $\sqrt{p(1-p)}$, also arbitrarily small. This poses some problems in using the volume of the QSE as a measure of correlations. With these states, examining the expression for the QSE volume, the correlations are non-zero such that the numerator in Eq.~\eqref{eqvol} is non-zero, but the maximisation of the expression works by minimising the denominator. This means that obtaining $b$ close to 1 causes the QSE to grow and accounts for the odd observation of boundary states with vanishing correlations.

For the last family of states, we use the states which maximise concurrence at set values of the linearised entropy~\cite{Munro,Steve}, which read 
\begin{equation}
\rho^{(conc)}= \left(\begin{array}{cccc}
g(\gamma) & 0 & 0 & \gamma/2\\
0&1-2g(\gamma)&0&0\\
0&0&0&0\\
\gamma/2& 0 & 0 & g(\gamma) \end{array}\right)
\end{equation}
with 
\begin{equation}
g(\gamma)=\begin{cases}
\gamma/2, &\gamma \geq2/3\\
1/3, &\gamma<2/3 \end{cases}.
\end{equation}
With these definitions, the family $\rho^{(R3)}$ that we are after is simply given by a linear mixture of $\rho^{(R2)}$ with the aforementioned states $\rho^{(conc)}$. That is
\begin{equation}
\rho^{(R3)}=m_3\rho^{(conc)}+ (1-m_3)\rho^{(R2)},
\end{equation}
with $m_3\in[0,1]$ the mixing parameter of the two families. 

In Fig.~\ref{figvol} we have generated $10^8$ random two-qubit states. The boundary states were identified with a mixture of analytical, and numerical techniques. Initially, various previously known families of states, such as the Maximally Entangled Mixed States (MEMS) were plotted, leading to the identification of $\rho^{(R1)}$ with  Werner states. The method of Lagrange Multipliers was used to maximize $V_A$, thus getting close to $\rho^{(R2)}$ and looking the actual boundary states by an extensive numerical investigation on the parameters in the family. A linear mixing of the previous two families was used as a first ansatz for such a class of states. In fact, this choice turned out to be not optimal, although the attempt yielded results very close to the border, but with one small region close to $\rho^{(R1)}$ left uncovered. This led us to believe that the linear mixing should include another family and not $\rho^{(R1)}$. As the Werner states behave very similarly to MEMS for concurrence they were chosen to replace $\rho^{(R1)}$ in the linear mixing and success was achieved. The evidence that these states lie on the boundary is purely numerical. Also shown in Fig~\ref{figvol} is the distribution for random Bell-diagonal states (i.e. states that are diagonal in the Bell-state basis), which are dubbed here as X-states in light of the form that their density matrix takes in the computational basis.

\subsection{Optimal families for the Steerable Weight}
\begin{figure}
	\begin{center}
		\includegraphics[width=1.\columnwidth]{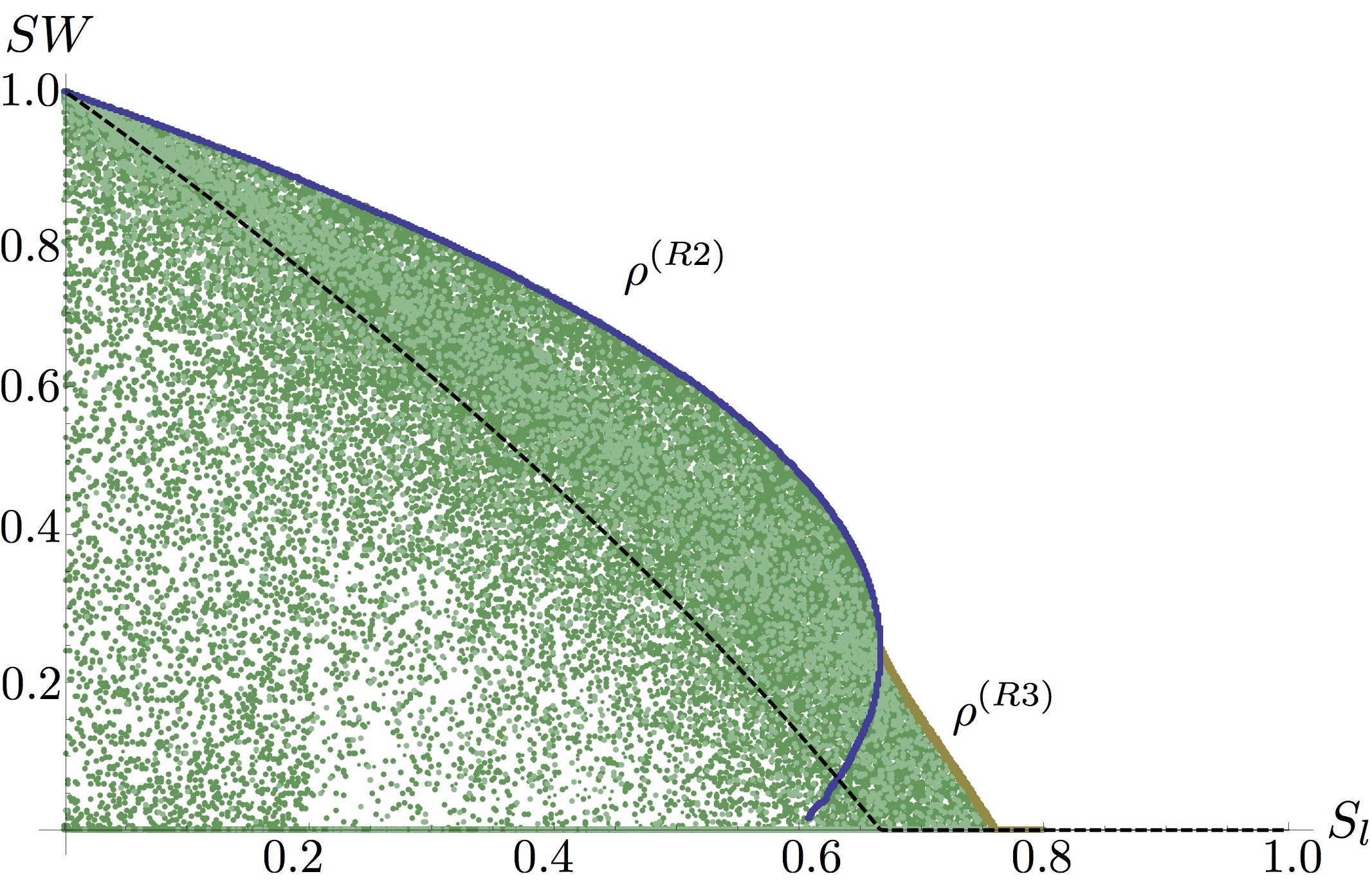}
		\caption{(Color online) 
			A graph of the SW versus the linear entropy for random states. The two families of states which maximise the SW are shown. The Werner states (black, dashed) are shown for the sake of comparison. The maximally discordant states ($\rho^{(R2)}$, blue, solid) maximise the SW up to a value of $S_L={2}/{3}$ for the linear entropy. The third family ($\rho^{(R3)}$, yellow points), which results from a linear mixture of the first two, is shown as a distribution.}
		\label{figWei}
	\end{center}
\end{figure}

The recently proposed Steerable Weight (SW)~\cite{Miguel} is a convex steering monotone~\cite{Gallego} which provides a quantitative answer to the following question: ``{\it Given an arbitrary assemblage, how well can we emulate it with a LHV-LHS description?}''. The problem is formulated in terms of a semi-definite program, which can be solved by convex optimisation. In this case the optimisation procedure tries to minimise the difference between the given assemblage and an arbitrary LHV-LHS. For an assemblage $\sigma_{b|B}$, the SW is defined as the minimum positive $\nu$ such that
\beq
\sigma_{b|B}=\nu\tilde{\rho}_{b|B}+(1-\nu)\sigma_\xi,
\eeq
where $\tilde{\rho}_{b|B}$ is an arbitrary assemblage and $\sigma_\xi$ corresponds to a local hidden state. For instance, in the case of a Bell state, one obtains $\nu=1$ due to the face that the LHV-LHS cannot reproduce $\sigma_{b|B}$ at all and instead the program has to build it out of $\tilde{\rho}_{b|B}$ entirely. The minimization over $\nu$ can be formulated in terms of the following semi-definite program (SDP). We aim at 
\beq
\label{SDP1}
\begin{aligned}
	&\text{find max}\,\text{Tr}\sum_{\xi}\sigma_\xi,\\
	&\text{such that}~\sigma_{b|B}-\sum_{\xi}D_\xi(b|B)\sigma_\xi\ge0~~\forall b,B\\
	&\text{with}~\sigma_\xi\geq0~~\forall \xi.
\end{aligned}
\eeq
Here $D_\xi(b|B)$ are the deterministic single-party conditional probability distributions. The SDP stated in Eq.~\eqref{SDP1} can be efficiently run using freely available numerical tools~\cite{CVX}. In what follows we will restrict ourselves to six assemblages per two-qubit system that we test. These correspond to the two possible outcomes for each of three measurements along the Pauli axis. Later on we will explore how the Pauli basis may not be optimal for trying to find the maximal amount of steering, but for our boundary states the Pauli axis is optimal for making measurements along and as such will not affect the discussion of our results.

The families of states which maximise the steerable weight turn out to be the same as the ones that maximise the volume. This is reported in Fig.~\ref{figWei}. However, at high linear entropy all correlations can be explained in terms of an LHV-LHS model and the steerability is zero. The region in which the Werner states would be expected to maximise the steering is already too highly mixed to observe steering, so we are left with just two families of states which maximise the steerable weight $\rho^{(R2)}$ and $\rho^{(R3)}$.
\begin{figure}[b]
	\begin{center}
	\includegraphics[width=0.9\columnwidth]{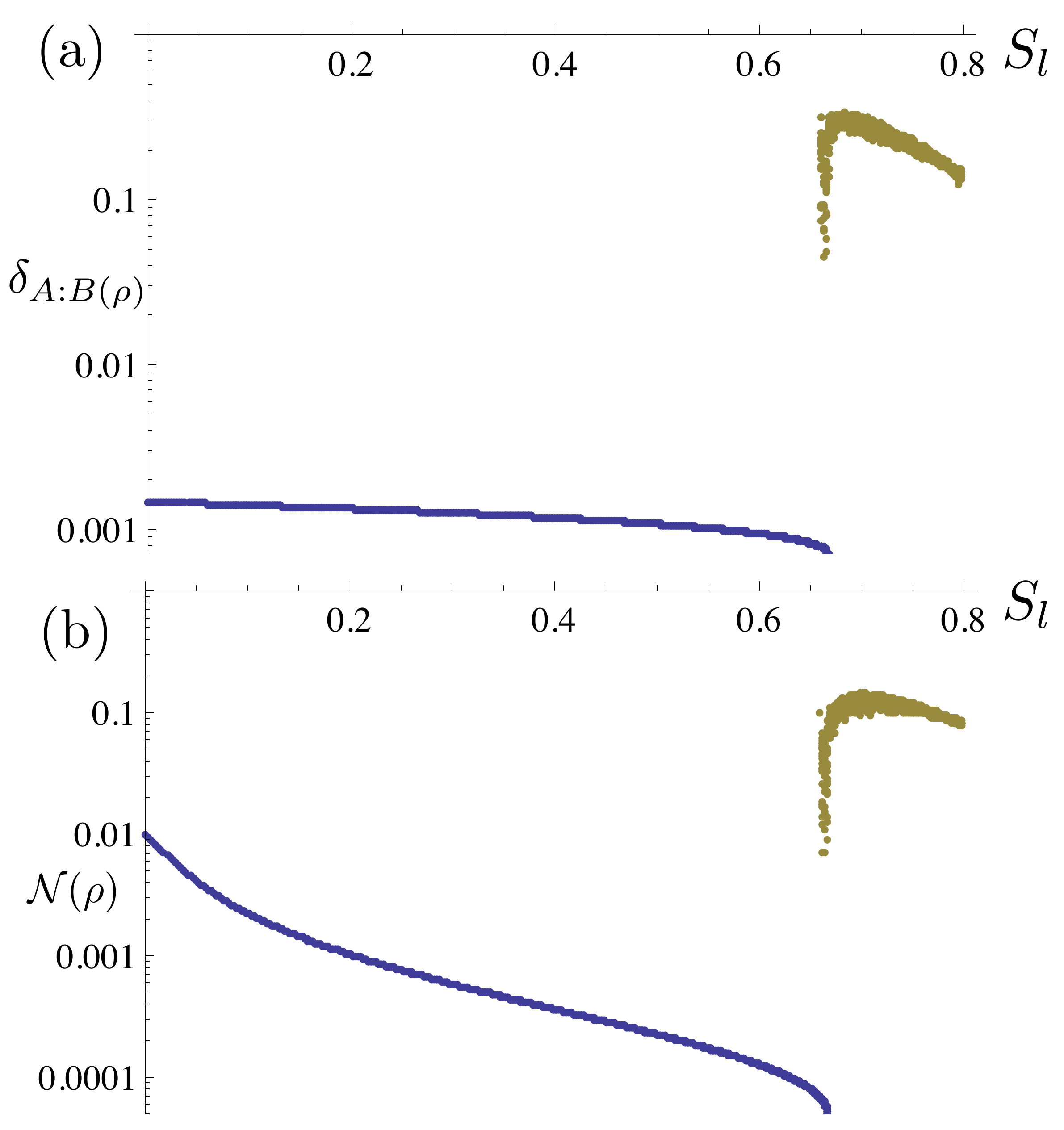}
		\caption{(Color online) 
			Graphs showing the (a) Quantum discord and (b) negativity for the two families which maximise the SW at a given linear entropy. The colours correspond to the same families as in the previous Figures. The correlations are lower at low linear entropy counter-intuitively to what one would expect.}
		\label{figDiscNeg}
	\end{center}
\end{figure}

The amount of correlations that are exhibited by the family $\rho^{(R2)}$ on the boundary are actually much lower than one would expect. In fact there are less correlations between the qubits (measured by mutual information, discord, entanglement) for $\rho^{(R2)}$ than $\rho^{(R3)}$. The result is shown in Fig.~\ref{figDiscNeg}. This is quite counter-intuitive given that the SW is higher for $\rho^{(R2)}$, however, this result follows naturally from the proof in Appendix E of Ref.~\cite{Miguel} that all entangled pure states are maximally steerable. Indeed, all we are doing is taking a pure entangled state and mixing it linearly with $\ket{01}\bra{01}$ to create our family. In this way we end up with states that have a very high SW but a very low amount of correlations for a given amount of linear entropy. The same states maximise the QSE volume for a given linear entropy, for the same reasons illustrated in Sec.~\ref{optimal}. 
We found the QSE volume to be 1 for all pure entangled states also. Since all pure entangled states give maximal QSE volume and SW it's a trivial thing to say that the MSMS are given by mixing these with a purely classical state until the appropriate amount of mixedness (linear entropy) is obtained.

\subsection{Optimal families for the Robustness of Steering}
The recently introduced robustness of steering (RS)~\cite{Piani} is another convex steering monotone~\cite{Gallego} holding the potential to capture the multifaceted features of steering. This measure follows previous measures of entanglement robustness and works by asking the question: ``{\it How much mixing must one add to a given assemblage in order for it to be fully described a LHV-LHS?}" For an assemblage $\sigma_{b|B}$, the RS is defined as the minimum positive $t$ such that
\beq
\sigma_{b|B}=(1+t)\sigma_{\xi}-t\tilde{\rho}_{b|B},
\eeq
 where $\tilde{\rho}_{b|B}$ is an arbitrary assemblage and $\sigma_{\xi}$ corresponds to a local hidden state. For instance, in the case of a Bell state (where the most mixing must be added) one finds a value of $t\simeq0.26$ whereas in the case of a product state no mixing need be added and one obtains $t=0$. This minimisation can be formulated as the following SDP
\beq
\begin{aligned}
&\text{find max}\,\text{Tr}\sum_{\xi}\sigma_\xi\\
	&\text{such that}~~\sum_{\xi}D_\xi(b|B)\sigma_\xi\geq\sigma_{b|B}~\forall b,B\\
	&\text{with}~\sigma_\xi\geq0~\forall \xi.
\end{aligned}
\eeq

\begin{figure}
\begin{center}
\includegraphics[width=1.\columnwidth]{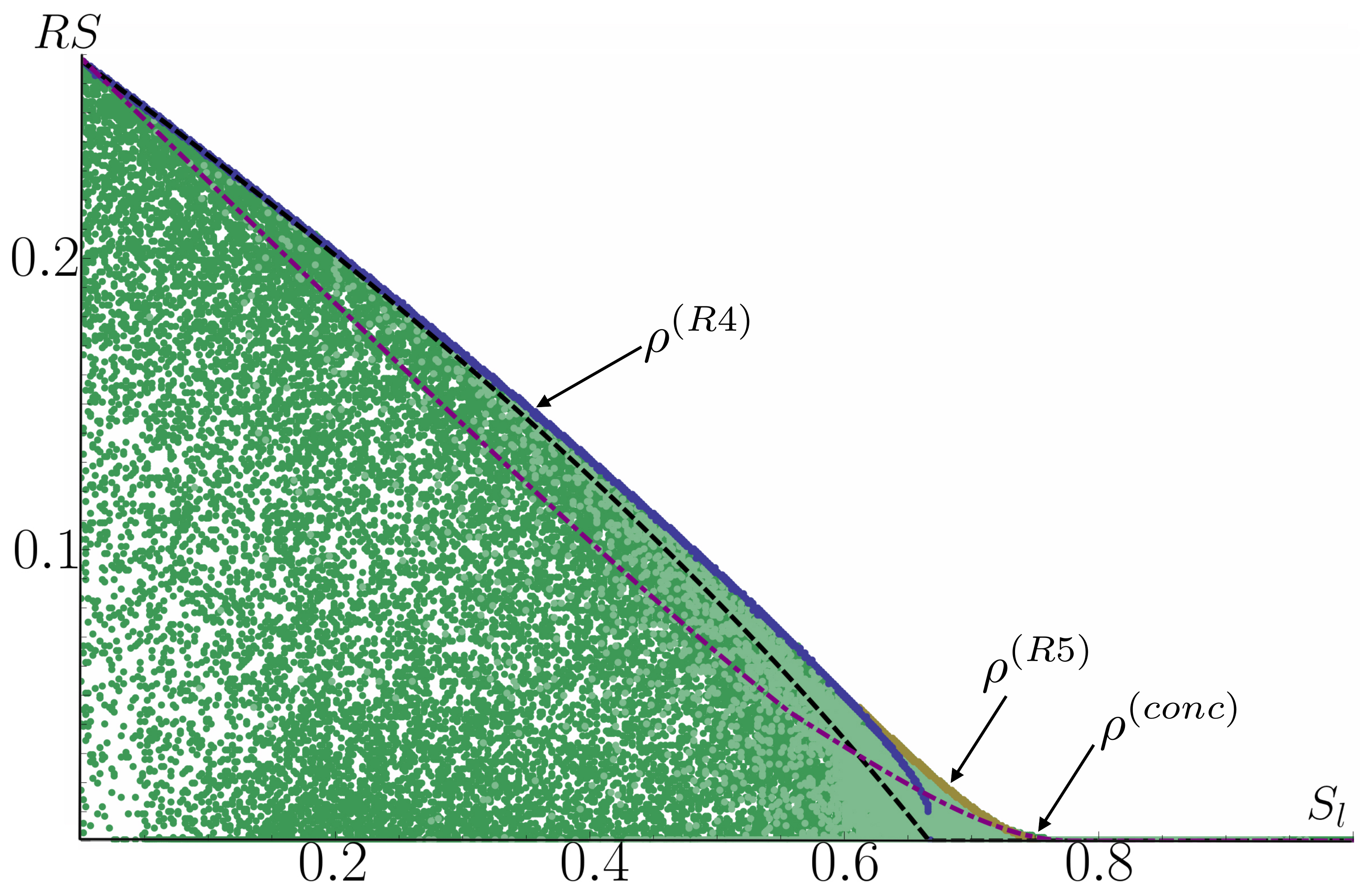}
\caption{(Color online) 
A graph of the RS versus the linear entropy for random states. The three families of states which maximise the RS are shown. The Werner states (black, dashed) are shown for the sake of comparison. The doubly asymmetric Bell states ($\rho^{(R4)}$, blue, solid) and the family that maximises the concurrence ($\rho^{(conc)}$, purple, dot-dashed). The third family ($\rho^{(R5)}$, yellow points), resulting from a linear mixture of the first two, is also shown.}
\label{figRob}
\end{center}
\end{figure}

There are three families of states that maximise the RS. The first is given by
 $$
 \rho^{(R4)}=m_4\ket{\tilde{\Phi}^+}\bra{\tilde{\Phi}^+}+(1-m_4)\ket{\tilde{\Phi}^-}\bra{\tilde{\Phi}^-}
 $$
 with $\ket{\tilde{\Phi}^-}=\sqrt{p}\ket{01}+\sqrt{1-p}\ket{10}$. The second family is simply given by the family of states which maximise the concurrence ($\rho^{(conc)}$) and the last family is given by the linear mixture of the previous two:
 $$
 \rho^{(R5)}=m_5\rho^{(R4)}+(1-m_5)\rho^{(conc)}.
 $$
 Unfortunately, as the method for finding these quantifiers of genuine EPR steering involves the convex optimisation of semidefinite programs, we were unable to use analytic techniques to identify the boundary states. Therefore, our analysis of both RS and SW involves mostly extensive numerical searches based on previously known MEMS. Our evidence that these lie on the boundary is purely numerical.
  
The RS measure introduced in Ref.~\cite{Piani} appears to be finer-grained than the SW, as it is not maximal for all pure entangled states. For instance the state $\sqrt{0.999}\ket{00}+\sqrt{0.001}\ket{11}$ has unit SW, but the amount of noise that needs to be added to this to have non-steerable assemblages is very small indeed. The RS for this state is very close to zero. In an analogous way the MSMS for the SW and QSE volume have zero RS.

\section{Optimal Measurement Basis}
\label{basis}

During the course of our investigation we discovered that the Pauli basis is not always the optimal measurement one --- namely, the one that maximises the effects of Alice's measurements on Bob's qubit. In order to fully explore the steering properties, one should prepare assemblages using every possible measurement for Alice. However, as there are infinitely many projective measurements, this is not a feasible pursuit.

\begin{figure}[t!]
\begin{center}
\includegraphics[width=1.\columnwidth]{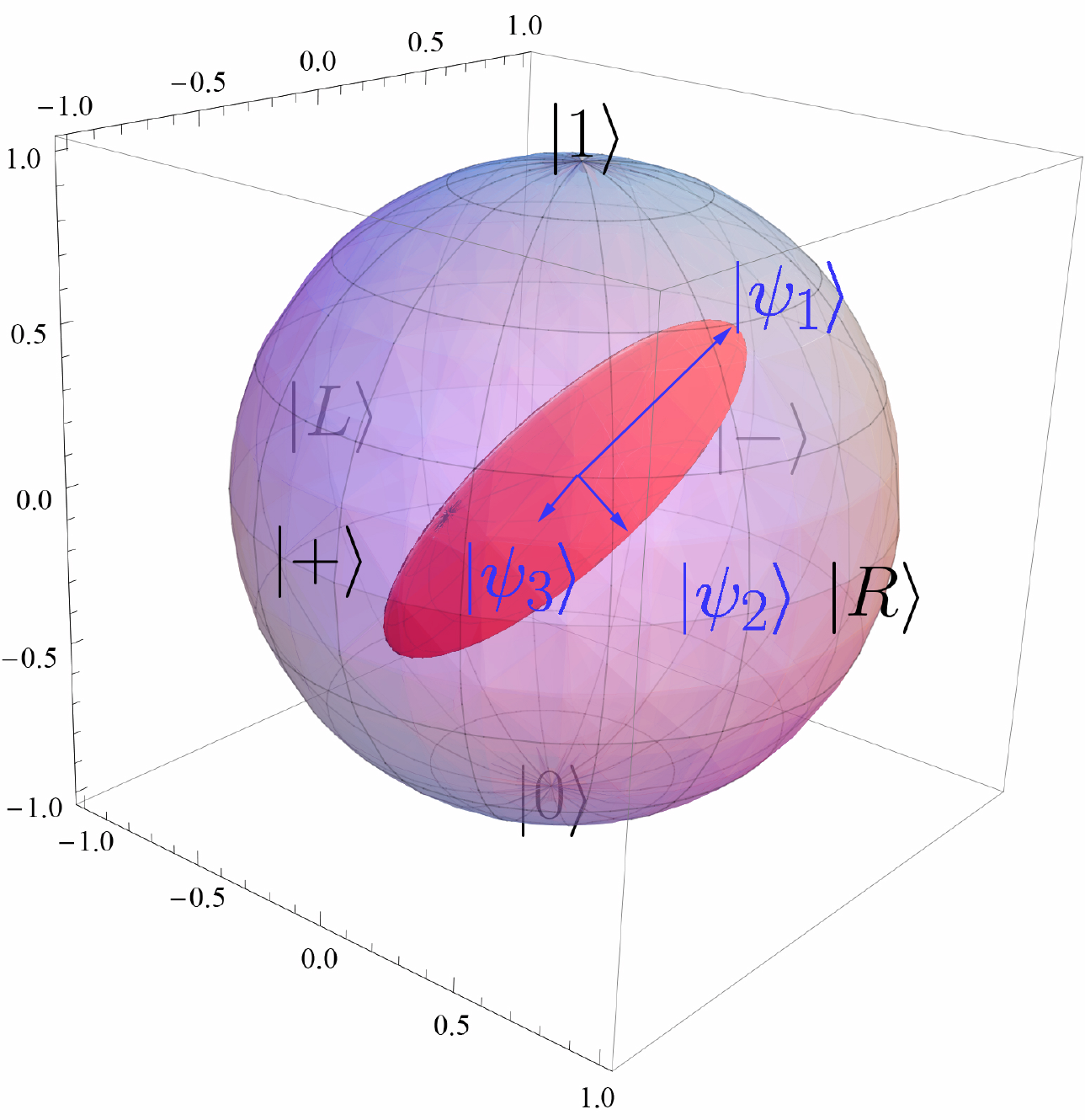}
\caption{(Color online) 
Diagram of a QSE, showing how measurement in the Pauli basis is suboptimal in the case of this two-qubit state. Using only three types of measurement the best we can do in terms of maximising our usage of QSE volume is to use the semiaxes of the QSE as the basis for our measurements.}
\label{figEli}
\end{center}
\end{figure}

\begin{figure}[b!]
\begin{center}
\includegraphics[width=0.9\columnwidth]{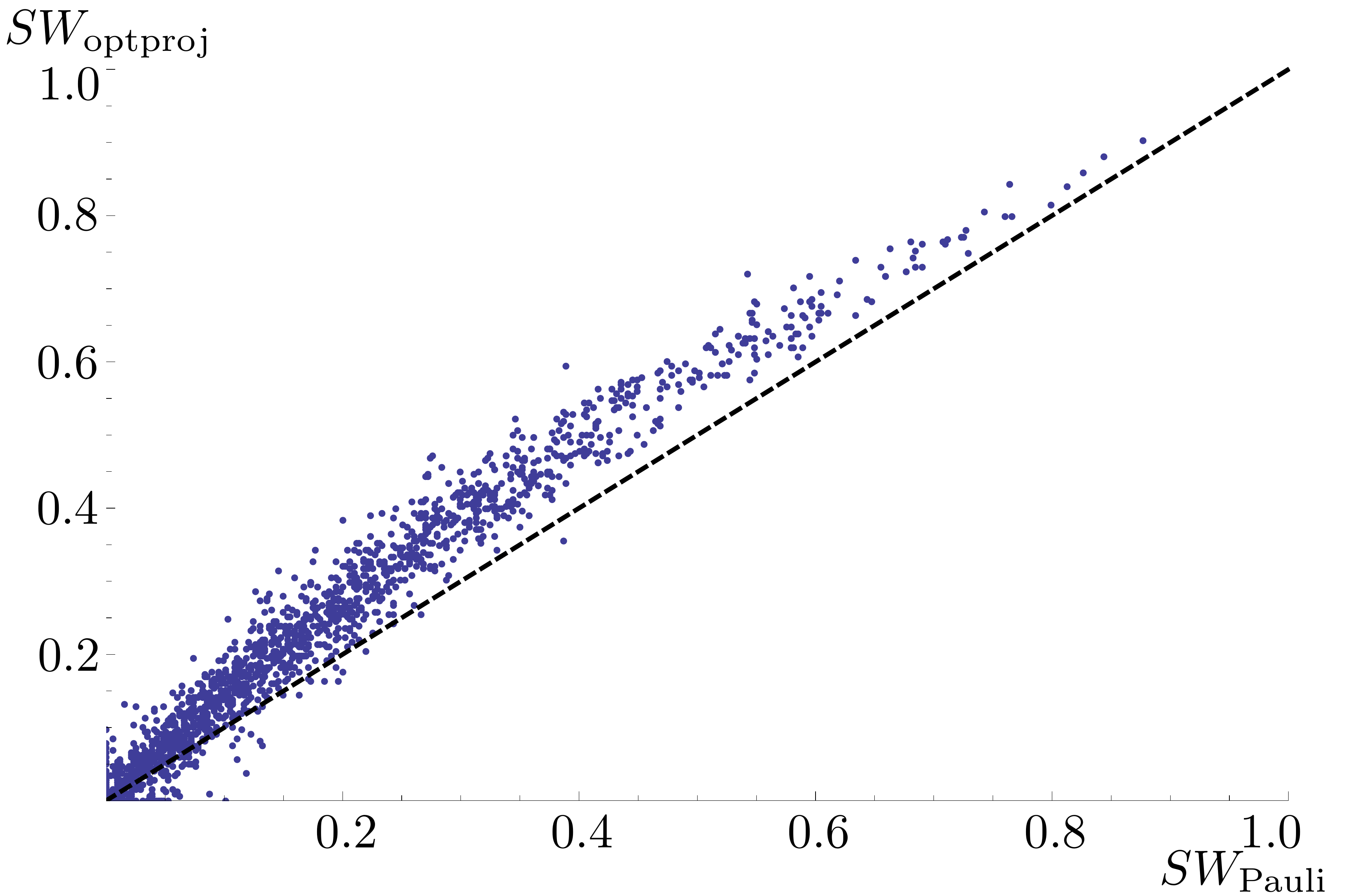}
\caption{(Color online) Diagram of a QSE, showing how measurement in the Pauli basis can be suboptimal. Using only three types of measurement the best we can do in terms of maximising our usage of QSE volume is to use the semiaxes of the QSE as the basis for our measurements.}
\label{figMeas}
\end{center}
\end{figure}

If we restrict ourselves to three measurements then we can look at the QSE as a guide to what basis our measurements should be in. If the ellipsoid is rotated at all from the Pauli basis then our three optimal orthogonal measurements should also be rotated. We can use the semiaxes of the QSE to work out the basis that we should choose for our projective measurements, or simply apply a unitary to the two-qubit state to rotate the QSE back to having its semiaxes correspond with the Pauli basis. Using this approach we find, in 98\% of cases, an improvement to the steerable weight obtained. This is reported in Fig.~\ref{figMeas}.

 The idea behind our scheme is to use the QSE to identify the optimal directions in which to measure on Bob to obtain the maximal amount of SW. The rationale behind our approach is based on the results that we have highlighted above, i.e. that QSE and SW are very much related, thus suggesting the use of the former to find a better estimate of the latter. The methodology highlighted below demonstrates that such an intuition is (very often) successful. In order to access the largest possible volume of our QSE we should be looking for the projectors which are directed along the semiaxes of the QSE. The semi-axes of an ellipsoid are given by the eigenvectors, ${\bf e}_v$ of the ellipsoid matrix, in our case given by ${\tilde T}{\tilde T}^T$. Then we reverse-engineer the projection directions ${\bf p}_B$ to be made on Bob so that Alice's post-measurement state is in the directions of ${\bf e}_v$ by solving the equation,
 $$
 \tilde T {\bf p}^x_B = {\bf e}^x_v,
 $$
where $x$ is an index describing which measurement direction we are looking at. In our case we are looking at three measurements so $x$ ranges from $1$ to $3$. Now we simply take these projector directions and use them to work out our basis using the Pauli matrices:
$$
\hat P^x_B={\bf p}^x_B\cdot\{\hat{\sigma_x},\hat{\sigma_y},\hat{\sigma_z}\},
$$
where $\hat P^x_B$ represents one of the bases which we now use to find the projectors for our measurement on Bob. Now we are left with the task of using the semi-definite program of Ref.~\cite{Miguel} to work out the steerability. In order to test our method we shall use the ``normal" method, i.e. use the Pauli basis for our measurements on randomly generated two-qubit states ($\rho$), and our method, i.e. use the $\hat P^x_B$ basis for our measurements on the transformed two-qubit states ($\rho'$). Our results are reported in Fig.~\ref{figMeas}.

We have found a remarkable improvement of the SW for the vast majority of the cases. In only 2\% of the cases, usually at low value of steerability, the transformation $\rho\rightarrow\rho'$ brings the state into a regime where correlations between the two qubits can be better explained classically, i.e. the new state is less steerable than the old one. This accounts for the values that lie below the line in the graph. In light of this we tentatively propose that we could better identify the steerability of a generic state as $SW=\text{max}\left(SW_{\text{Pauli}},SW_{\text{optproj}}\right)$.

\section{Conclusions}
\label{conc}

We have examined a measure of two-qubit correlations (the QSE volume) and two measures of genuine EPR steerability, namely the WS and the RS, identifying the sets of two-qubit states that maximize the ``steering" at a given degree of mixedness for all of the measures considered. For values of linear entropy greater than 0.76 there is no state with positive steerability for either measure. Interestingly, below this value, the families of states that maximize the QSE volume and the SW turn out to coincide. This gives strong evidence of the close links between these two quantities, despite the fact that the former is not a genuine measure of EPR steerability.

We have found that the third measure, the Robustness of Steering, has instead different families of maximal states: the first family is a mixture of Bell states, the second one is composed of known MEMS states,
and the third one is a mixture of the latter two. We have attributed the difference between these families and the ones that maximize the QSE volume and the SW to the fact that the latter are generally not resilient against noise.


Moreover, we have shown that there exists a method, using the QSE, to better calculate the steerability. In particular, such method provides better projectors to use in place of the Pauli ones, and is successful in $\sim98\%$ of cases. It should be noted, though, that the Pauli basis is indeed optimal for all of the above analysis on the maximally steerable mixed states. 

Our results suggest that the ties between the QSE and genuine measures of steering are tighter then previously noticed, being more then just qualitative in nature. In addition, the MSMS
states that we have introduced can be the most useful for one-sided device independent cryptography at a given amount of noise, given the known links between the latter task and steering. Finally, the introduction of both MSMS and the optimal projectors to reveal their properties could be useful for future experimental studies aiming at generating steering.


\acknowledgements
We thank O. G{\"u}hne and Marco Piani for helpful discussions and acknowledge support from DEL, the EU FP7-funded Collaborative  Project TherMiQ, the John Templeton Foundation (grant number 43467), the Julian Schwinger Foundation (grant number JSF-14-7-0000), and the UK EPSRC (grants EP/M003019/1 and EP/P00282X/1).

\end{document}